\documentclass[twocolumn, a4paper,usenatbib]{aastex62}
\usepackage{ae,aecompl,graphicx,amsmath,amssymb,bm,array,booktabs} 
\usepackage{blindtext, mathtools}
\usepackage[T1]{fontenc}

\newcommand{\ie}{i.e.,~}
\newcommand{\eg}{e.g.,~}

\begin{document}

\title{Optimal neutron-star mass ranges to constrain the equation of
  state of nuclear matter with electromagnetic and gravitational-wave
  observations}

\author{L. R. Weih}
\affiliation{Institut f\"ur Theoretische Physik, Goethe Universit\"at
Frankfurt am Main, Germany}
\author{E. R. Most}
\affiliation{Institut f\"ur Theoretische Physik, Goethe Universit\"at
Frankfurt am Main, Germany}
\author{L. Rezzolla}
\affiliation{Institut f\"ur Theoretische Physik, Goethe Universit\"at
  Frankfurt am Main, Germany}

\begin{abstract} 
Exploiting a very large library of physically plausible equations of
state (EOSs) containing more than $10^{7}$ members and yielding more than
$10^{9}$ stellar models, we conduct a survey of the impact that a
neutron-star radius measurement via electromagnetic observations can have
on the EOS of nuclear matter. Such measurements are soon to be expected
from the ongoing \textit{Neutron Star Interior Composition Explorer}
mission and will complement the constraints on the
EOS from gravitational-wave detections. Thanks to the large statistical
range of our EOS library, we can obtain a first quantitative estimate of
the commonly made assumption that the high-density part of the EOS is
best constrained when measuring the radius of the most massive, albeit
rare, neutron stars with masses $M\gtrsim2.1\,M_\odot$. At the same time,
we find that radius measurements of neutron stars with masses
$M\simeq1.7-1.85\,M_\odot$ can provide the strongest constraints on the
low-density part of the EOS. Finally, we quantify how radius measurements
by future missions can further improve our understanding of the EOS of
matter at nuclear densities.
\end{abstract}


\section{Introduction}
\label{sec:intro}
%
\begin{figure*}
\begin{center}
  \includegraphics[width=0.85\textwidth]{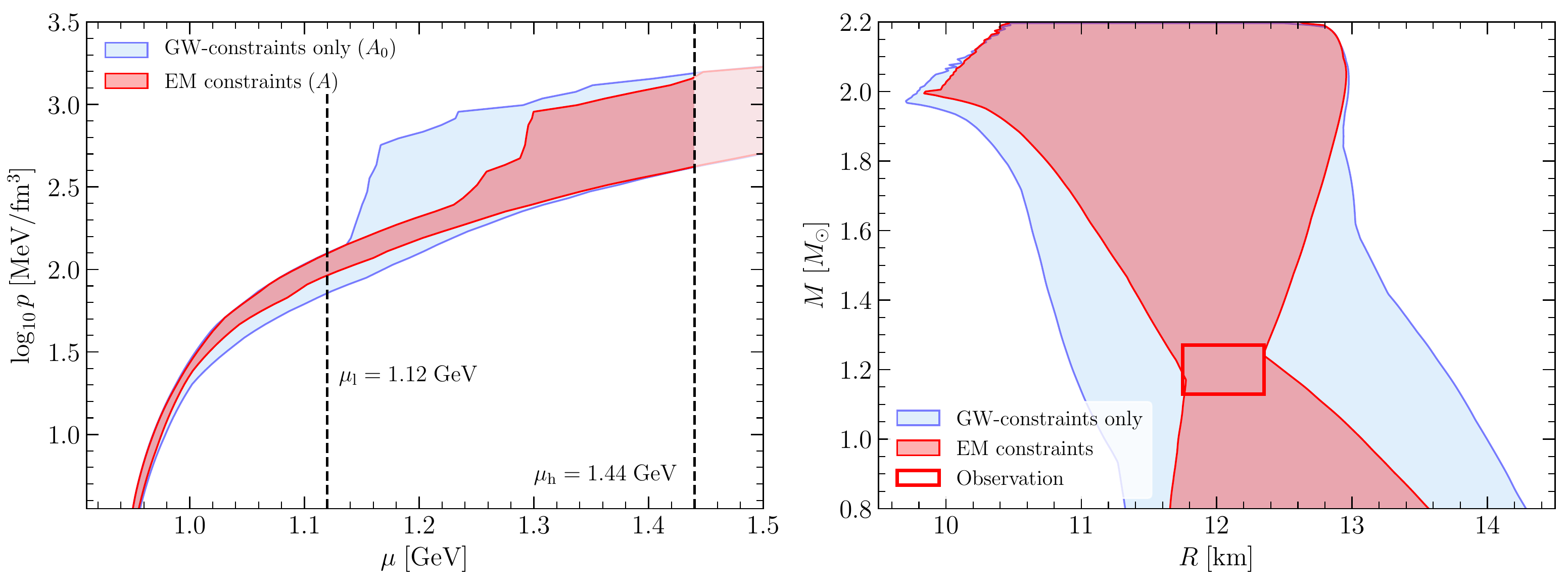}
  \caption{\textit{Left:} areas in $(p,\mu)$ plane spanned by all EOSs
    fulfilling the GW constraints (light blue, $A_0$) and by EOSs that
    also fulfill a constraint as expected from an electromagnetic
    measurement by NICER (red, $A$). The dashed vertical lines limit the
    different ranges of the EOS in terms of the chemical potential
    $\mu$. \textit{Right:} the same areas but in the $(M,R)$ plane. The
    red box marks a representative measurement by NICER. An animated
	version is available \href{https://doi.org/10.5281/zenodo.3363615}{online}
	\citep{Animation}. The animation 
    explains our method as described in Sec. \ref{sec:methods} graphically.
	It first shows how the EOSs span the areas in $(p,\mu)$ and $(M,R)$ and then
        uses an examplatory radius measurement to illustrate the relation between
	the DOC, this measurement, and the corresponding constrained area. Finally, the
	animation shows how the DOC is computed for all possible radius measurements and
	how a color-coded representation of DOCs as the ones in Fig. \ref{fig:fig3}
	is generated.
  \label{fig:fig1}}
\end{center}
\end{figure*}%

The recent detection of gravitational waves (GW) from the inspiral of two
neutron stars, \ie GW170817 \citep{Abbott2017_etal}, in combination with
the observation of its electromagnetic counterpart \citep{Abbott2017b}
has already led to a number of astrophysical constraints on the nuclear
equation of state \citep[EOS;][]{Bauswein2017b, Margalit2017, Paschalidis2017,
Shibata2017c, Abbott2018b, Annala2017, Burgio2018, De2018, Fattoyev2017, 
Lim2018, Malik2018, Montana2018, Most2018,Radice2017b,Raithel2018,Rezzolla2017,
Ruiz2017,Tews2018, Gill2019, Koeppel2019, Lim2019, Tews2019}. In
the near future, these constraints are expected to be complemented by
precise measurements of neutron-star masses and radii via electromagnetic
observations yielding constraints similar to that of a distant GW event
\citep{McNeil_Forbes2019}. The \textit{Neutron Star Interior Composition
  Explorer} (NICER) \citep{Arzoumanian2014,Gendreau2016} is set to
deliver these constraints via X-ray modeling \citep[see][for a
  review]{Watts2016}. Anticipating these results, considerable work has
been dedicated toward inferring the EOS from electromagnetic
observations via Bayesian inference and different parametrizations for
the EOS \citep{Ozel2009,Read:2009a,Steiner2010,Raithel2017,Miller2019}.

In this work we combine the GW constraints with possible mass and radius
observations as those expected from NICER. A similar study that combines
constraints has recently been undertaken by \citet{Fasano2019}, who, in
addition to the GW constraints, also used in their Bayesian analysis
radius measurements from thermonuclear bursts in accreting neutron stars
\citep[see also][for earlier studies]{Guillot2013,Lattimer2014,Ozel2015}.

We here assess how much the already constrained EOS will be further
restricted by precise measurements of the neutron-star radius.
\citet{Greif2019} have recently shown that inferring properties of dense
matter via Bayesian analysis from any astrophysical observations, \ie
from GW constraints or constraints on the neutron-star mass and radius,
depends sensitively on the prior assumptions of the underlying EOS
parameterization.  In view of this, we do so not by relying on a Bayesian
analysis or any approach involving probabilities. Rather, we consider a
very large sample of EOSs whose completeness allows us to span the whole
range of physically plausible EOSs given a generic and conservative
parameterization. Using much smaller samples, similar approaches have
already been employed in the past,
\citep[\eg][]{Hebeler2013,Kurkela2014,Annala2017}, and allow for general
statements that are valid independently of prior assumptions. This is
true as long as the EOS sample is sufficiently large to provided
convergent results as in the case of our EOS library (see also the Appendix). 
In this way, by considering a large set of possible
outcomes of the NICER mission, we are able to assess the impact of future
radius observations on our ability to further constrain the EOS and to
specify the most promising mass ranges to be targeted for deriving the
tightest constraints.

\section{Methods}
\label{sec:methods}

In order to relate possible neutron-star radius measurements to the EOS,
we consider equilibrium neutron-star models that are solutions to the
Tolman-Oppenheimer-Volkoff (TOV) equations constructed from a
comprehensive set of different EOSs obeying recent constraints from GW
and kilonova observations \citep{Most2018}. These EOSs are constructed
following the approach of \citet{Most2018} \citep[see also][for a similar
  approach]{Kurkela2014}, by combining state-of-the art calculations from
chiral effective field theory (CEFT) \citep{Drischler2016,Drischler2017},
describing the behavior of nuclear matter accurately close to
nuclear-saturation density, together with results from perturbative QCD
(pQCD) calculations \citep{Kurkela2010,Fraga2014} for the high-density
regime. We model the remaining region typically found in the core of
neutron stars by parameterizing the EOS with piecewise polytropes. We
only allow for EOSs that are causal, \ie for which the speed of sound,
$c_s$, is smaller than the speed of light, and are able to support a mass
of at least $1.97\,M_\odot$ \citep{Antoniadis2013}. By matching to the
pQCD limit, all of our EOSs automatically fulfill the criterion that
asymptotically the sound speed is given by $c_s=\sqrt{1/3}$.

In this way, we have constructed a set of $\simeq2.5\times10^7$
physically plausible EOSs from which we have computed
$\simeq3.8\times10^{9}$ stellar models. All of our EOSs are purely
hadronic and hence do not account for a first-order phase
transition. While this is a limitation that we will address in a future
work, stellar models with phase transitions have been found to represent
a small fraction of the physically plausible EOSs \citep{Most2018}. To
this set, which we refer to as the \textit{``complete''} set, we apply
the constraints derived from the GW event GW170817
\citep{Abbott2017_etal,Abbott2018b}. These constraints have been derived
in several works and can be summarized as follows:

\smallskip
\textit{(i)} The maximum mass reached by a sequence of nonrotating
neutron stars does not exceed $\simeq2.2\,M_\odot$
\citep{Margalit2017,Shibata2017c,Rezzolla2017,Ruiz2017}.  This is a
conservative value for the maximum mass that agrees with the error bounds
of the four papers listed above. We also compare our results to the case
of a higher value for this upper limit, \ie $\simeq2.3\,M_\odot$, as
suggested in \citet{Shibata2019} and in agreement with the upper
bound of \citet{Rezzolla2017}. A discussion in the Appendix
will illustrate how a different choice of the maximum
mass impacts our results only marginally.

\textit{(ii)} The tidal deformability of a $1.4\,M_\odot$ star,
$\Lambda_{1.4}$, is constrained to be in the range $290<\Lambda_{1.4}<580$, 
where the upper limit is the observational constraint from GW170817
\citep{Abbott2018b}, while the lower limit is deduced from the analysis
of \citet{Most2018} and is consistent with a number of other studies
\citep{Abbott2018b,Coughlin2018,De2018,Kiuchi2019}.

\smallskip
Imposing these constraints reduces our set of EOSs considerably, leaving
us with $\approx20\,\%$ of the complete set. We refer to this reduced
library as the \textit{``constraint-satisfying''} set and it is marked
with the light blue area in Fig.~\ref{fig:fig1}, whose left panel
provides a representation in the pressure-chemical potential plane,
$(p,\mu)$, while the right panel in the mass-radius plane $(M,R)$.

A number of remarks are useful at this point. First, it is essential that
the sample of EOSs considered is sufficiently large so as to provide a
robust statistical representation after the GW constraints are
imposed. We have verified that this is the case by considering sets of
$2.5$, $1.5$, and $1.0\times10^7$ EOSs, which all provide the same
statistical results. Second, the evidence that the GW constraints remove
about $80\,\%$ of the physically plausible EOSs, underlines the
importance of a sufficiently large sample of EOSs that allows for
convergent results (see the discussion in the Appendix). Third, several
works have also derived constraints on the radius of a $1.4\,M_\odot$
star
\citep{Annala2017,Burgio2018,De2018,Malik2018,Most2018,Raithel2018,Tews2018};
these constraints are compatible with our range of admissible tidal
deformabilities. Fourth, the range for $\Lambda_{1.4}$ is also in
agreement with constraints stemming from the threshold mass to prompt
collapse \citep{Bauswein2017b,Koeppel2019}.  Finally, \citet{Gamba2019}
have recently shown that while the constraints on the radius might
sensitively depend on the description of the EOS for the neutron-star
crust, the constraints on $\Lambda_{1.4}$ are more robust.

To quantify how well potential observations of neutron-star radii can
further constrain the EOS in addition to the constraints already imposed,
we need to define a measure of the degree to which an EOS is
constrained. More specifically, because the pressure $p(\mu)$ is a
thermodynamical potential that fully describes the EOS at $T=0$, we
consider as a reference measure of the properties of our EOS library the
area spanned by the constraint-satisfying set of EOSs in our sample, \ie
\begin{equation}
\label{eq:area}
A_{0}\coloneqq\!\!\int\!\! d \mu\!\!\int\!\! d p(\mu) =
\!\!\int_0^{\mu_{\rm{h}}}\!\!\!\!
{\left[\mathrm{max}(p(\mu))-\mathrm{min}(p(\mu))\right] d\mu}\,,
\end{equation}
where the maximum and minimum in the integral is taken at each point
$\mu$ and over the set of constraint-satisfying EOSs. We choose $\mu_{\rm
  {h}}=1.44\,\rm{GeV}$ which is the largest chemical potential found in
our constraint-satisfying sample and reflects the fact that neutron stars
cannot probe the EOS beyond the largest densities found in their cores
(the highest number density in our sample is $\sim6.5\,n_0$, where
$n_0=0.16\,\rm{fm}^{-3}$ is the nuclear-saturation density). Note that by
varying the range in the chemical potential over which the integral in
Eq. \eqref{eq:area} is performed we can probe in a differential manner
the low-density and the high-density regimes of the EOS. In particular,
we will define the low (high)-density reference areas $A_{0,\mathrm{l}}$
($A_{0,\mathrm{h}}$) if the integral \eqref{eq:area} is instead performed
in the range $[0,\mu_{\rm{l}}]$ ($[\mu_{\rm{l}},\mu_{\rm{h}}]$), where
$\mu_{\rm{l}}=1.12\,\rm{GeV}$\footnote{This value corresponds roughly to
  where we match the EOS from CEFT. Since the matching is done at a fixed
  number density, it corresponds to different values of $\mu$ for each
  EOS.} (see black dashed lines in Fig. \ref{fig:fig1}).

\begin{figure*}[t]
  \centering
  \includegraphics[width=0.75\textwidth]{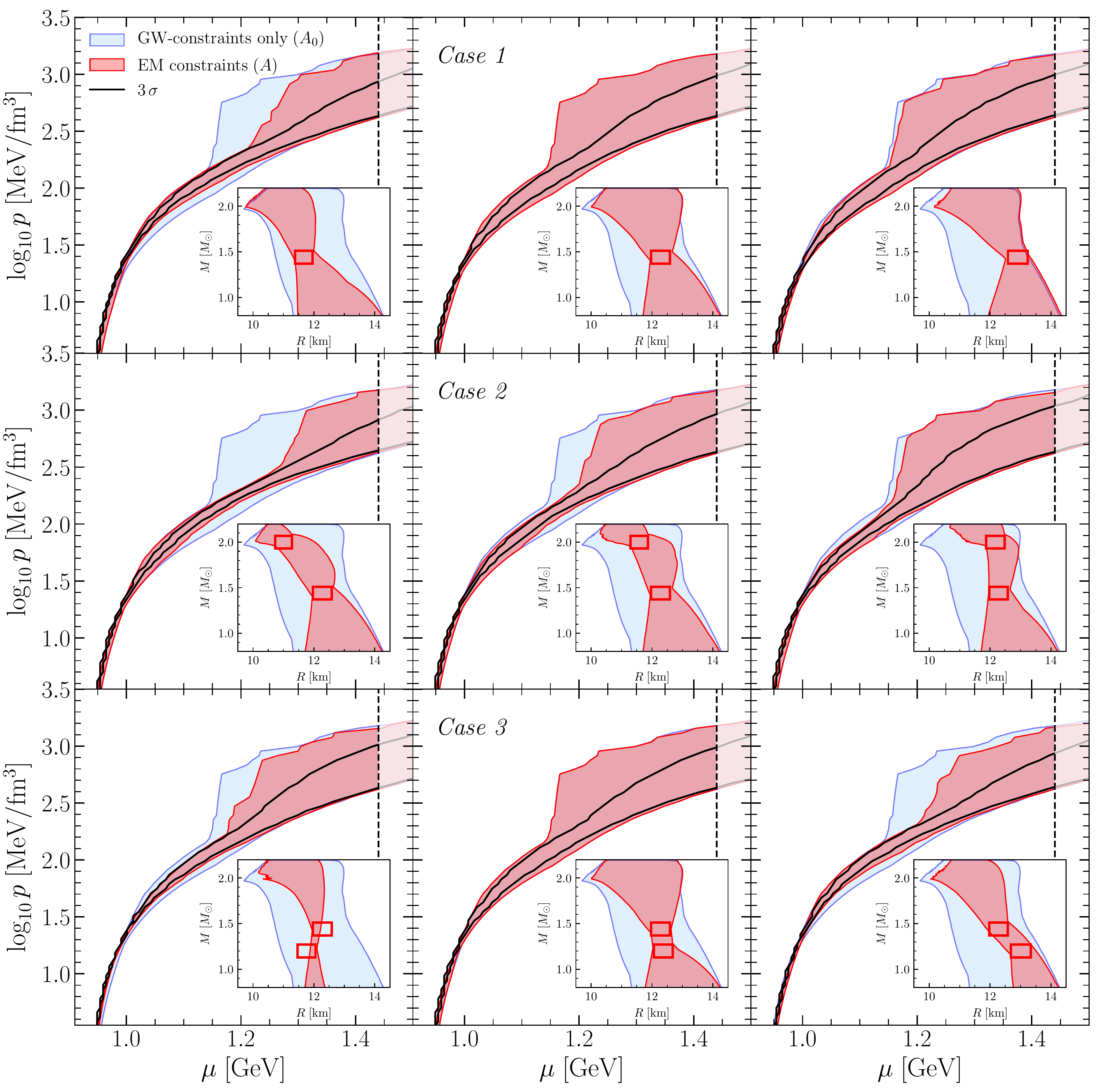}
  \caption{Same as in Fig. \ref{fig:fig1} but specialized to the
    reference \textit{Case 1}-\textit{Case 3} in the text. For each row
    the central column refers to radius measurements inferred by
    \cite{Most2018}, while the left/right columns explore the cases when
    the radius is $10\%$ smaller/larger. Shown with solid black lines in
    each panel are the $3\,\sigma$ confidence levels obtained from a
    Bayesian analysis of the EOS distributions.}
  \label{fig:fig2}
\end{figure*}

Having defined a reference area $A_0$, we can ask how it changes when we
impose an additional constraint from a hypothetical mass and radius
measurement. For simplicity we represent such an observation in terms of
a square box in the $(M,R)$ plane whose boundaries are determined by the
errors of the measurement. Any EOS not passing through this box can then
be discarded as being in disagreement with the observation. An
illustration of this logic is shown in Fig. \ref{fig:fig1}, where the new
area $A$ is represented by the red-shaded area and where a hypothetical
measurement has been made of a neutron star with mass $1.2\pm0.07\,M_\odot$ 
and radius $12.0\pm0.3\,{\rm{km}}$ (red box in
Fig. \ref{fig:fig1}). In this way, we can measure the \textit{``degree of
  constraint''} (DOC) of the observation simply as
\begin{equation}
\label{eq:chi}
	\chi\coloneqq1-\frac{A}{A_{0}}\,,
\end{equation}
where $\chi\ll1$ refers to a measurement that is not constraining the EOS
appreciably, while $\chi\simeq1$ would indicate a measurement that can
set significant constraints on the EOS, as when the red-shaded area in
Fig. \ref{fig:fig1} shrinks to a line. In logical analogy with the
definition of \eqref{eq:chi}, we can measure the DOCs for the low
(high)-density region of the EOS, $\chi_{\rm{l}}$ ($\chi_{\rm{h}})$, by
simply replacing $A$ and $A_0$ with the corresponding areas computed when
$\mu\in[0,\mu_{\rm{l}}]$ ($[\mu_{\rm{l}}, \mu_{\rm{h}}]$).

Using these definitions, we can now assess how well a given mass/radius
measurement would constrain the EOS and, more importantly, which part of
the EOS is actually constrained. Before doing that, however, we underline
that the DOC, $\chi$, depends on the prior with which the EOSs are 
constructed. As long as the EOS sample size is large enough, however, 
all DOCs computed with any prior will converge to the same value.
This is an important added value of our approach compared with other works.

\section{Results}
\label{sec:results}

Hereafter, we will consider three main cases that are believed to be
possible outcomes of the NICER mission \citep{Arzoumanian2014,Watts2016},
namely \citep[see also][for similar choices]{Greif2019}:

\smallskip
\textit{Case 1:} Only one of the primary targets, \ie PSR J0437-4715,
with a mass of $1.44\,M_\odot$ will be successfully measured yielding a
radius of $12.28\pm0.31\,\rm{km}$.

\textit{Case 2:} Both primary targets, \ie PSR J0437-4715 and PSR
J0030+0451 will be successfully measured. For the latter the mass is
unknown and we here consider a massive star with $2.0\,M_\odot$ with the
same error as for PSR J0437-4715. For the radius we assume
$11.58\pm0.29\,\rm{km}$.

\textit{Case 3:} The same as \textit{Case 2}, but with $1.2\,M_\odot$ for
PSR J0030+0451 and a radius of $12.37\pm0.31\,\rm{km}$.

\smallskip
\noindent The radii in the cases above are taken as the most likely
values reported by \citet{Most2018}, together with a $5\,\%$ relative
error on the radius \citep{Arzoumanian2014} and a fixed uncertainty of
$\pm 0.07\,M_\odot$ for the mass\footnote{For a larger error of $\pm
  0.14\,M_\odot$, or $10\,\%$ uncertainty in the radius measurement, we
  obtain essentially the same results, with slightly smaller values for
  the DOCs.}, where the latter is motivated by the typical accuracy of
known neutron-star masses \citep{Ozel2016}.

\begin{figure*}
\begin{center}
  \includegraphics[width=0.75\textwidth]{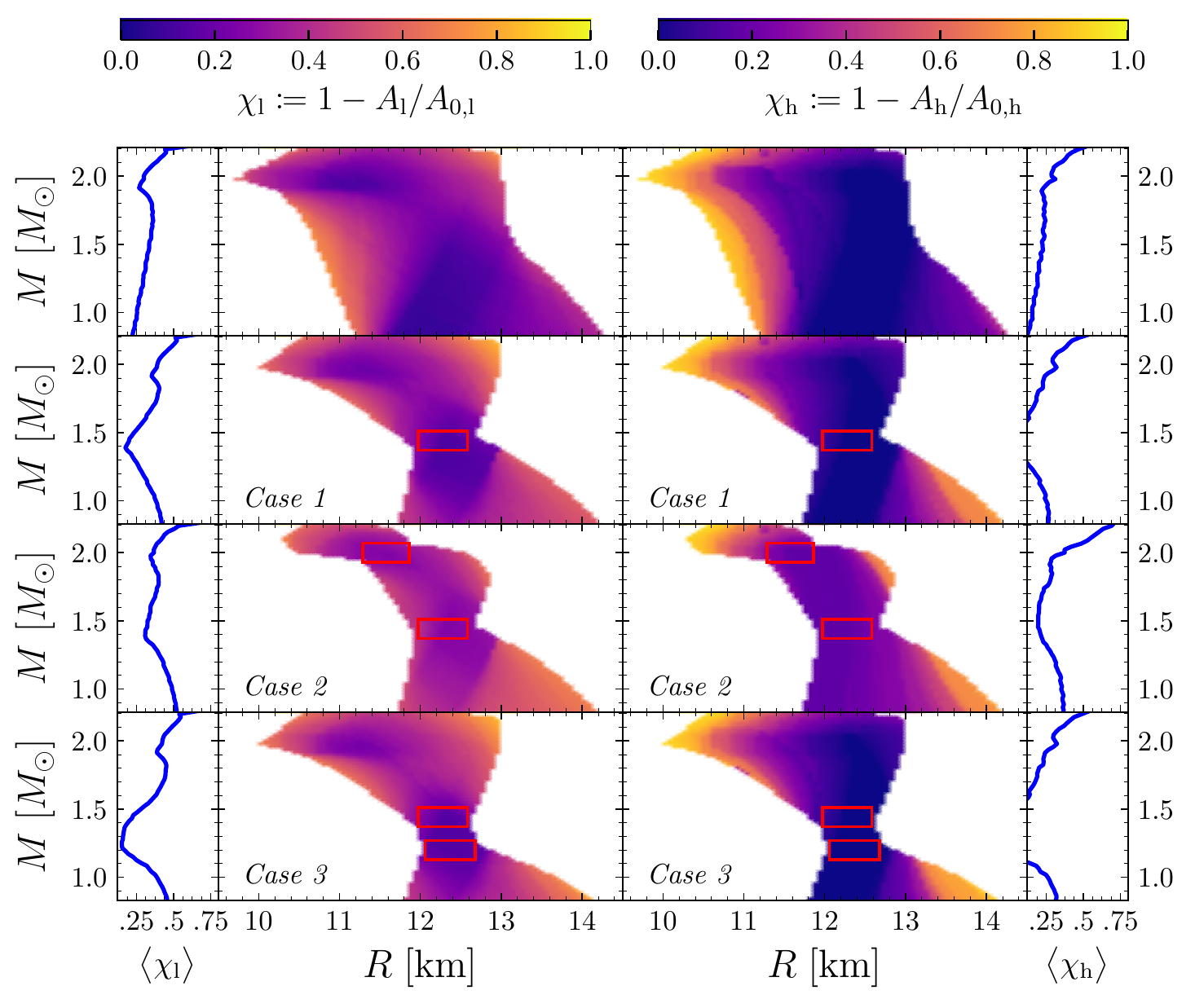}
  \caption{Degrees of constraint (DOCs) in the $(M,R)$ plane for the low-
	(left panels, $\chi_{\rm{l}}$) and high-density (right panels,
	$\chi_{\rm{h}}$) regions of the EOS and for all possible hypothetical
    observations. We assume the constraint of such an observation in
    addition to already existing constraints. The top row refers to GW
    constraints only, while the other rows refer to an assumed outcome of
    the NICER mission (red boxes), \ie \textit{Case 1}-\textit{Case 3},
    respectively. The left and right side panels show the radius-averaged
	DOCs, $\langle\chi_{\rm{l,h}}\rangle$.}
        \label{fig:fig3}
\end{center}
\end{figure*}

Shown in the middle panel of the first row of Fig. \ref{fig:fig2} is
a graphical representation of how the DOC is defined, \ie as the ratio of the
red to the blue area. This panel shows these areas for
\textit{Case 1} (shown in the
  inset are the corresponding areas in the $(M,R)$ plane). Note that in
this case $\chi\simeq0$ as the blue- and red-shaded areas essentially
coincide, thus indicating that a measurement of this type would constrain
the EOS only marginally. Similar considerations apply also when examining
a measured radius that is $10\,\%$ larger, as indicated in the top right
panel of Fig. \ref{fig:fig2}. On the other hand, the top left panel,
which refers to a measured radius that is $10\,\%$ smaller, shows that
$\chi\sim0.67$, thus indicating that measuring a compact star would
provide important constraints on the EOS. Stated differently, given a
$1.44\,M_\odot$ star, the most significant constraints on the EOS would
be obtained if the star had a radius $R_{1.44}\lesssim12\,{\rm{km}}$.

\begin{figure*}
\begin{center}
  \includegraphics[width=0.85\textwidth]{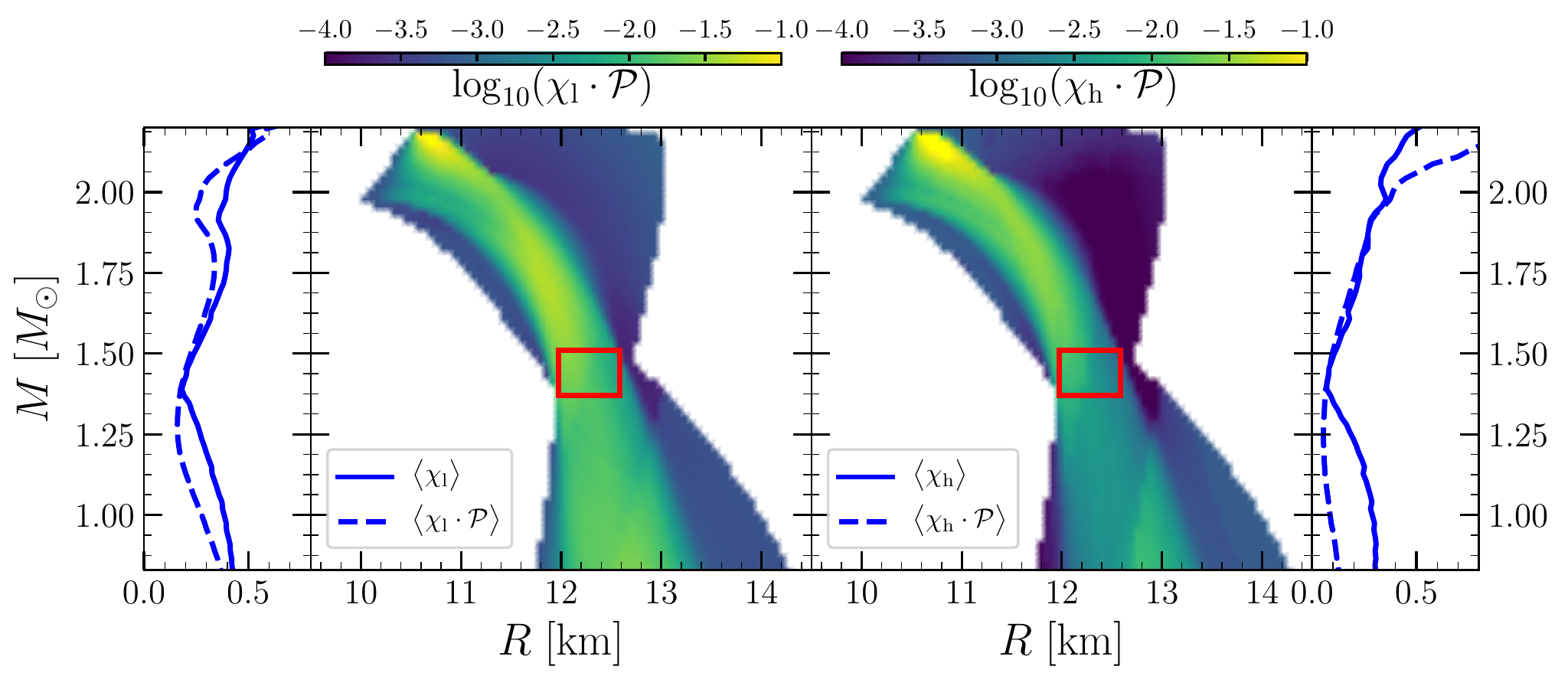}
  \caption{Same as in the second row of Fig. \ref{fig:fig3}
    (\textit{Case 1}), but showing the probability-weighted DOCs,
	$\chi_{\rm{l,h}}\cdot\mathcal{P}$. The left and right side panels
    show the corresponding deconvolved and radius-averaged DOCs,
	$\langle\chi_{\rm{l,h}}\cdot\mathcal{P}\rangle$.}
        \label{fig:fig4}
\end{center}
\end{figure*}

Similar conclusions follow also for \textit{Cases 2} and \textit{3},
which are reported, respectively, in the central panel of rows two and
three of Fig. \ref{fig:fig2}, while the corresponding left/right panels
refer to the same cases, but assuming a smaller/larger radius for PSR
J0030+0451. As a reference, we also show with black solid lines the
$3\,\sigma$ confidence levels obtained by computing the probability
distribution function (PDF) via a Bayesian analysis where we use the
maximum of a bivariate Gaussian distribution within the red box in the
$(M,R)$ plane for the likelihoods \citep[as done in, \eg][]{Raithel2017},
and the PDF obtained by \citet{Most2018} as prior. The area encompassed
by these lines shows that even when $\chi\simeq0$, constraints on the EOS
may still be obtained within a given confidence level. In particular, for
the high-density part of the EOS the $3\,\sigma$ confidence intervals
remain unchanged under various radius measurements. Because the precise
measure of these constraints will depend upon the choice of prior used
for the Bayesian analysis \citep{Greif2019}, we will here use the DOC
defined in Eq. \eqref{eq:chi}, which is free of this bias as it only
compares the outer limits of the EOS space covered, for which we can show
convergence irrespective of the prior.

The first row of Fig. \ref{fig:fig3} shows with a colorcode for each
point in the $(M,R)$ plane the DOC when an observational box is centered
around that point. In other words, we consider the DOC for all possible
mass/radius measurements. Note that we now distinguish the DOCs relative
to the low- (left panel) and high-density (right panel) regions of the
EOS. Obviously, an observation providing a box at the edges of the area
spanned in the $(M,R)$ plane would provide the best constraints as the
underlying EOSs there are the scarcest. The left and right side columns
of Fig. \ref{fig:fig3} show with blue lines the averaged DOCs
$\langle\chi_{\rm{l,h}}\rangle$, namely, the radial averages of $\chi_{\rm{l}}$ 
and $\chi_{\rm{h}}$ taken at a fixed mass. It is then evident that
the best constraints on the EOS can be expected from measuring the radius
of a neutron star that is as massive as possible, because these stars have
densities in their cores for which the EOS is least known. Interestingly,
Fig. \ref{fig:fig3} also shows that there is a local maximum for
$\langle\chi_{\rm{l}}\rangle$ around $M \sim1.8\,M_\odot$, thus
suggesting that -- apart from very massive and rare stars with 
$M\gtrsim2.1\,M_\odot$ -- a radius measurement in the range of
$1.7-1.85\,M_{\odot}$ would yield the best constraints on the low-density
regime of the EOS. While this is true only on average and the DOC will
depend on the exact measured value, this is the first time that such an
evidence is found. We also remark that if going to higher masses yields
smaller DOCs, this is most likely due to the presence of a lower limit on
the maximum mass $M_{_{\rm{TOV}}}>1.97M_\odot$.

The last three rows of Fig. \ref{fig:fig3} show the same DOCs for
\textit{Case 1}-\textit{3} and hence describe the degree to which the EOS
is constrained for an additional radius measurement beyond NICER. In all
three cases, the local maximum around $\sim1.8\,M_\odot$ is again
present for $\langle\chi_{\rm{l}}\rangle$, underlining the importance that
a star in this mass range would have on constraining the low-density
regime of the EOS.

The averaged DOCs $\langle\chi_{\rm{l,h}}\rangle$ seem to suggest
that the EOS would be considerably constrained also for a radius
measurement of a star with $M\lesssim1.1\,M_\odot$, thus making low-mass
stars almost as promising as the high-mass ones. Such large averaged DOCs
are the results of the increase in $\chi_{\rm{h}}$ following the
measurement of a low-mass star with a large radius (\ie $\gtrsim13.5\,{\rm km}$;
see the orange part of the right panels for $\chi_{\rm{h}}$ in Fig. 
\ref{fig:fig3}). In practice, however, such large DOCs do
not take into account that in our library a large radius for such low-mass
stars is rare and hence unlikely to be observed. This can be countered by
convolving the DOC with the PDF $\mathcal{P}$ of encountering a star with
a given mass and radius. While a trivial step to take given that we
already have this information \citep{Most2018}, 
it also introduces a new direct dependency on the prior used in
building the set of EOSs, and would sacrifice the general validity 
of our results.

Notwithstanding this, we report in Fig. \ref{fig:fig4} the result of such
weighting and show the PDF-weighted DOCs, $\chi_{\rm{l,h}}\cdot\mathcal{P}$, 
and the averaged PDF-weighted DOCs,
$\langle\chi_{\rm{l,h}}\cdot\mathcal{P}\rangle$, for \textit{Case 1}, with
the PDF here being the same as that used for the computation of the
$3\,\sigma$ lines in Fig. \ref{fig:fig2}. In this way, we learn that
while large DOCs $\langle\chi_{\rm{l,h}}\rangle$ for a $1.8\,M_\odot$ star
are robust, this is not the case for a low-mass star with mass
$\lesssim1.1\,M_\odot$. This is particularly evident for the high-density
part of the EOS, for which 
$\langle\chi_{\rm{h}}\cdot\mathcal{P}\rangle\ll\langle\chi_{\rm{h}}\rangle$. 
We find this behavior in all scenarios:
while a large $\chi_{\rm{l}}$ is robust and indicative of a genuine high
DOC, a large $\chi_{\rm{h}}$ at low masses may need to be weighted by the
probability of finding stars in this mass range with such extreme radii.

\section{Conclusions}
\label{sec:conclusions}

We have presented an extensive survey of how electromagnetic measurements
of masses and radii of neutron stars, when combined with the constraints
from GW170817, will provide information on the EOS of nuclear matter. Our
findings confirm the expectation that radius measurements of stars with
masses $M\gtrsim2.0\,M_\odot$ represent the most effective manner of
setting constraints on the highest-density regimes of the EOS. At the
same time, our analysis reveals the existence of an optimal mass range of
$M\sim1.7-1.85\,M_{\odot}$, which would yield the best constraints on
the low-density regime of the EOS with observations beyond those made by
the NICER mission. Finally, low-mass stars with $M\lesssim1.1\,M_\odot$,
could also provide significant constraints on the low-density regime of the EOS.

The library of EOSs and stellar models presented here and providing a
convergent and complete coverage of physically plausible EOSs, is
publicly available
\href{https://doi.org/10.5281/zenodo.3260991}{online}
\footnote{\href{https://doi.org/10.5281/zenodo.3260991}{10.5281/zenodo.3260991}} \citep{EOSs}
and can be used as new GW detections of merging neutron-star binaries or
mass/radius measurements become available. Such detections can be
employed to provide even more stringent constraints on the properties of
the EOS of nuclear matter. The library will also be updated to
incorporate phase transitions and new observational constraints, such as
those on the maximum mass via the binary system PSR J0740+6620
\citep{Cromartie2019}.

\section*{Acknowledgements}

It is a pleasure to thank Cole Miller, Cecilia Chirenti and Anna Watts
for useful discussions. Support comes also in part from HGS-HIRe for
FAIR; the LOEWE-Program in HIC for FAIR; ``PHAROS'', COST Action CA16214
European Union's Horizon 2020 Research and Innovation Programme (grant
671698) (call FETHPC-1-2014, project ExaHyPE); the ERC Synergy Grant
``BlackHoleCam: Imaging the Event Horizon of Black Holes'' (grant
No. 610058).The calculations were performed on the GOETHE cluster at CSC
in Frankfurt.

\appendix
\label{sec:appendix}

In this Appendix we provide additional information on some of the issues
discussed in the main text.

\begin{figure*}[h!]
\begin{center}
  \includegraphics[width=1\textwidth]{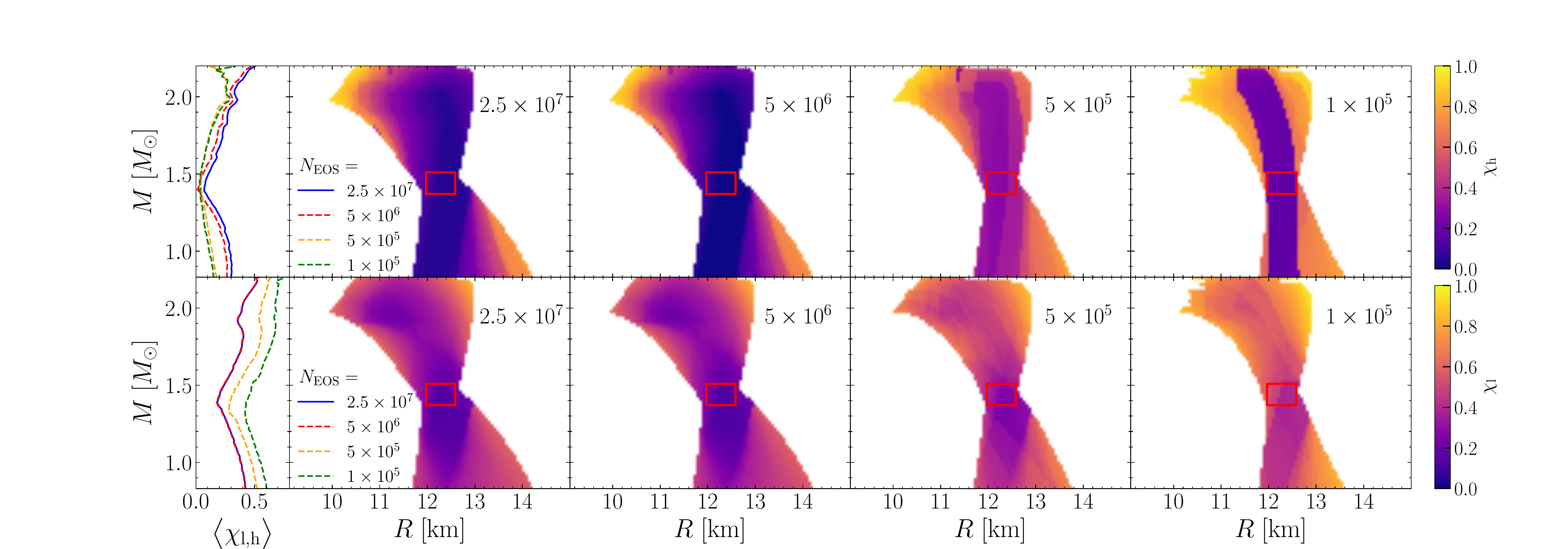}
  \caption{Same as the second row in Fig. \ref{fig:fig3}, but for
	different sizes of our EOS sample. From left to right we compare the
	results for samples of size $2.5\times 10^7$ (blue), $5\times 10^6$
	(red), $5\times 10^5$ (yellow), and $1\times 10^5$ (green).}
        \label{fig:fig5}
\end{center}
\end{figure*}

We start with providing convincing evidence that our results have been
obtained with a sufficiently large sample of EOSs to be statistically
robust and significant. In particular, the convergence of our results
is reported in Fig. \ref{fig:fig5}, which highlights the importance of
having a large enough sample of EOSs for reaching convergent
results. With smaller samples it is of course still possible to obtain
statistical results via, \eg Bayesian analysis.  For results over the
whole range of physically plausible EOSs, however, we find that at least
$\sim5\times10^6$ EOSs are necessary for our parameterization to populate
also the rare EOSs at the very stiff and soft ends of the spectrum. This
is mostly necessary because the largest part of our sample, \ie $\sim
80\%$ of it, is excluded after applying the GW constraints and it is
further reduced when applying the constraints from a hypothetical radius
measurement.

\begin{figure*}[h!]
\begin{center}
  \includegraphics[width=0.85\textwidth]{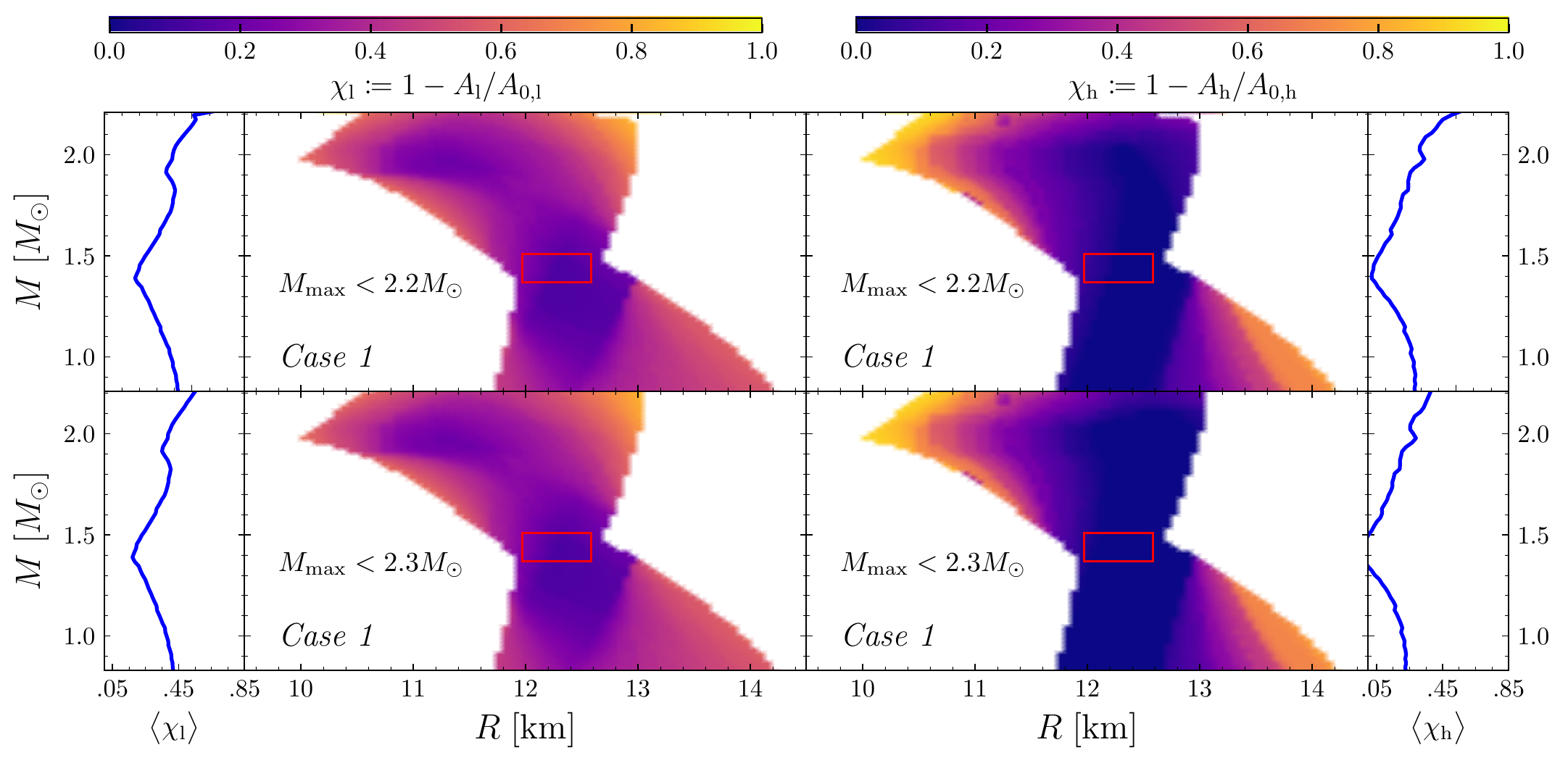}
  \caption{Same as the second row in Fig. \ref{fig:fig3}, but when
    using different values for the upper limit on $M_{_{\rm{TOV}}}$. The
    top and bottom panels show the results for
    $M_{_{\rm{TOV}}}<2.2\,M_\odot$ and $M_{_{\rm{TOV}}}<2.3\,M_\odot$,
    respectively. }
        \label{fig:fig6}
\end{center}
\end{figure*}

Next, we show that our results are robust with respect to the precise
value chosen for the upper limit on the maximum mass
$M_{_{\rm{TOV}}}$. Although the value of $M_{_{\rm{TOV}}} \lesssim
2.2\,M_\odot$ is in agreement with a number of studies each using
different methods
\citep{Margalit2017,Shibata2017c,Rezzolla2017,Ruiz2017}, it has recently
been argued that this upper limit might be higher and that it is weakly
constrained to be $M_{_{\rm{TOV}}} \lesssim 2.3\,M_\odot$
\citep{Shibata2019} which is consistent with the upper uncertainty bound
provided by \citet{Rezzolla2017}, namely, $M_{_{\rm{TOV}}}
\lesssim 2.16^{+0.17}_{-0.15}\,M_\odot$. To assess the validity of our
results we report in Fig. \ref{fig:fig6}
the changes in the DOCs when considering this higher value.  As it can be
deduced clearly when comparing the results in the top
($M_{_{\rm{TOV}}}<2.2\,M_\odot$) and bottom panels
($M_{_{\rm{TOV}}}<2.3\,M_\odot$) of Fig. \ref{fig:fig6}, the impact of
considering a larger maximum mass is only marginal. More importantly,
the evidence that the mass range around $M\sim1.8\,M_\odot$ provides the
optimal constraint for $\langle\chi_{\rm{l}}\rangle$ remains unchanged.

\newpage


\end{document}